\documentclass{aa}  
\usepackage{multirow}
\usepackage{graphicx}
\usepackage{txfonts}
\usepackage{subcaption}         
\usepackage{placeins}           
\newcommand{\kpc}{\,\mathrm{kpc}}

\begin{document}

\title{Reconciling Galactic rotation curve constraints with stellar stream modeling}

    \author{Mingyu Chang\inst{1}\corrauth{mingyu.chang@obspm.fr}
        \and François Hammer\inst{1}
        \and Yanbin Yang\inst{1}
        }
    
    \institute{LIRA, Observatoire de Paris, Université PSL, CNRS, Place Jules Janssen, 92195 Meudon, France}
    
    \date{Received 20 June 2026 / Accepted 18 August 2026}

\abstract
  {Recent \textit{Gaia}-based measurements of the Milky Way rotation curve and stellar-stream modeling give significantly different estimates of the Galactic dynamical mass beyond Galactocentric radii of $15~\kpc$.  The stream-based model predicts an outer halo five times more massive than that predicted by the \textit{Gaia} rotation curve.}
  {We aim to test the impact of analytic assumptions used in stream modeling and to assess whether the currently available stream constraints can distinguish between low- and high-mass Galactic potentials. }
  {We first compared globular-cluster disruption in analytic and N-body Milky Way potentials. We then modeled Palomar~5 and ATLAS--Aliqa Uma, which are unique in probing the outer region beyond $R_{\rm GC}\sim15~\kpc$ and are the most relevant to understanding the mass discrepancy. Both streams have usable constraints on sky position, proper motion, line-of-sight velocity, and RR~Lyrae distance. They were modeled for both a rotation-curve-based and a stream-based Galactic potential.}
  {In the N-body simulations, tidal shocks have a stronger effect on the closer orbit than on the more distant orbit, and therefore do not naturally explain the outer-Galaxy mass discrepancy. For the streams Palomar 5 and ATLAS–Aliqa Uma, simulations performed for both low- and high-mass Galactic potentials provide comparably good fits of their morphologies and kinematics. Neither potential provides a uniformly better match to all observables, and their differences are comparable to the present observational and modeling uncertainties.}
  {Current stream data do not discriminate between the low- and high-mass Milky Way models over the radial range probed by Palomar~5 and ATLAS--Aliqa Uma. This resolves the apparent tension between the rotation-curve- and stream-based constraints over this radial range.
  }

\keywords{Galaxy: kinematics and dynamics --
          Galaxy: structure --
          stellar streams -- 
          methods: numerical}

   \maketitle

\nolinenumbers

\section{Introduction}

The \textit{Gaia} mission \citep{Gaia2016} has greatly improved our view of Galactic dynamics by providing precise astrometry for nearly two billion stars. Yet recent \textit{Gaia}-based studies have not converged toward a single picture of the Milky Way mass distribution. In particular, rotation-curve and stellar-stream analyses give different results in the outer Galaxy, where the dark halo dominates and the available constraints become much sparser.

Since the discovery of flat rotation curves in spiral galaxies in the 1970s \citep{Rubin1978,Bosma1978}, rotation curves have remained one of the most direct probes of the dark matter distribution in disk galaxies. For the Milky Way, the combination of \textit{Gaia} astrometry with improved distance estimates has made it possible to measure the rotation curve over a much wider radial range than before. Using \textit{Gaia} Data Release (DR)
2 \citep{Gaia2018} proper motions and spectrophotometric distances, \citet{Eilers2019} found the first clear evidence of a declining Milky Way rotation curve out to $\sim 25$~kpc. This finding has since been supported by several independent studies based on \textit{Gaia} DR3 \citep{Gaia2023,Wang2023,Jiao2023,Ou2024}. At the same time, rotation-curve constraints are not assumption-free. In most modern analyses, the circular velocity is inferred through Jeans modeling, which requires the disk tracer population to be treated as approximately axisymmetric and close to dynamical equilibrium. These assumptions become increasingly difficult to justify in the outer disk, where warps, flares, bending modes, and other large-scale perturbations can be important. In addition, the inferred rotation curve depends on the adopted tracer density profile \citep{Koop2024} and on how systematic uncertainties are treated.

Stellar streams provide a complementary way to constrain the Galactic potential. In the hierarchical formation picture, low-mass systems such as globular clusters and dwarf galaxies are progressively disrupted by tidal forces while orbiting in the host halo, releasing stars into leading and trailing tidal tails \citep{Combes1999}. Because the debris remains dynamically coherent for long periods, stellar streams retain memory of the gravitational field through which they evolved and are therefore sensitive tracers of the Galactic mass distribution, especially in the halo where disk tracers are sparse. Using the \textsc{STREAMFINDER} atlas of \textit{Gaia} DR3 streams, \citet{Ibata2024} constructed a global Milky Way mass model by fitting the streams in an analytic Galactic potential, and obtained a rotation curve that deviates significantly from that of \citet{Eilers2019} beyond $R_{\rm GC} \sim 15$~kpc (see Fig.~\ref{fig:vc}). However, stellar streams are also not straightforward tracers of the potential. A stream does not exactly follow a single orbit, because stars are stripped with a finite spread in energies and angular momenta \citep{Sanders2013a,Sanders2013b}. Even in a smooth and static potential, tidal tails can develop epicyclic overdensities and other internal structures \citep{Kupper2008}. In the real Milky Way, stream formation can be further affected by time-dependent perturbations from baryonic structures, massive satellites, and the Galaxy's nonequilibrium history. Such effects are absent or only approximately represented in the smooth, static analytic potentials commonly adopted in stream modeling, and this can affect both the resulting stream structure and the inferred Galactic potential \citep{Bonaca2014,Buist2015,Sanderson2017}.

The tension between rotation-curve- and stream-based constraints may therefore stem from the modeling assumptions behind each method, systematic uncertainties in the observational data, or physical complexity not captured by current frameworks. In this work, we examine this issue from the stream side through analytic-potential tests and direct stream-modeling simulations.

First, we performed analytic-potential tests to examine whether the common approximation of a fixed analytic Galactic potential significantly affects globular-cluster disruption and the resulting stream morphology. We then focused on Palomar~5 (Pal~5) and ATLAS--Aliqa Uma (AAU), the two most distant thin streams in the \citet{Ibata2024} catalog that also have relatively good observational constraints. For these systems, we revisited the distance constraints using RR~Lyrae (RRL) stars and carried out stream-modeling simulations in both rotation-curve- and stream-based Milky Way potentials. The simulated streams were then compared directly with the available observations.

This paper is organized as follows. Section~2 describes the adopted Milky Way potential models, the stream sample, and the simulation setup. Section~3 presents the results of analytic-potential tests, assessing whether tidal-shock effects can alter stream morphology
and thus influence stream-based mass constraints. Section~4 presents the RRL distance constraints and the stream-modeling comparison for Pal~5 and AAU. Section~5 summarizes our main results and discusses their implications for the tension between rotation-curve- and stream-based Milky Way mass constraints.

\section{Methods}
\subsection{Galactic potential models}

\begin{figure}
    \centering
    \includegraphics[width=\columnwidth]{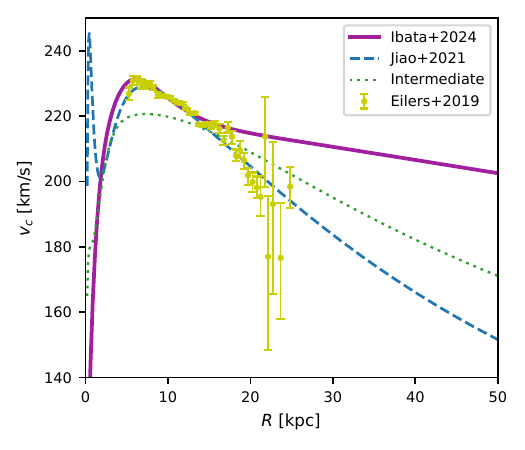}
    \caption{Circular velocity curves of the adopted Milky Way potential models. The dashed line shows the best-fit model of \citet{Jiao2021}, the solid line the stellar-stream best-fit model of \citet{Ibata2024}, and the dotted line the intermediate-mass model used for the tidal-shock tests. The points with error bars are the rotation-curve measurements of \citet{Eilers2019}.
    }
    \label{fig:vc}
\end{figure}

\begin{table*}
\centering
\caption{
Baryonic components of the Milky Way potential models adopted in this work. 
}
\begin{tabular}{llll}
\hline
Model & Component & Mass [$M_\odot$] & Scale parameters [kpc] \\
\hline

\multirow{4}{*}{I}
& Bulge & $1.067\times10^{10}$ 
& $b=0.3$ \\
& Thin disk & $3.944\times10^{10}$ 
& $a=5.3$, $b=0.25$ \\
& Thick disk & $3.944\times10^{10}$ 
& $a=2.6$, $b=0.8$ \\
\hline

\multirow{6}{*}{II}
& Bulge & $8.9\times10^9$ 
& $r_0=0.075$, $r_{\rm cut}=2.1$ \\
& H$_{\rm I}$ disk & $1.1\times10^{10}$ 
& $R_{\rm d}=7.0$, $R_{\rm m}=4.0$, $z_{\rm d}=0.085$ \\
& H$_2$ disk & $1.2\times10^9$ 
& $R_{\rm d}=1.5$, $R_{\rm m}=12.0$, $z_{\rm d}=0.045$ \\
& Thin stellar disk & $3.34\times10^{10}$ 
& $h_R=2.241$, $h_z=0.348$ \\
& Thick stellar disk & $0.84\times10^{10}$ 
& $h_R=1.74$, $h_z=0.858$ \\
\hline

\multirow{3}{*}{III}
& Bulge & $1.12\times10^{10}$ 
& $a=1.25$ \\
& Disk & $3.58\times10^{10}$ 
& $a=2.4$, $b=0.24$ \\
\hline

\end{tabular}

\label{tab:mw_potentials}
\end{table*}

The Milky Way is commonly modeled as comprising a central bulge, a stellar disk, a gas disk, and a dark matter halo. Different studies adopt different mass profiles for these components. In this work, to examine how stellar stream evolution depends on the underlying Galactic potential, we considered two Milky Way mass models inferred from two different methods: the rotation-curve-based model of \citet{Jiao2021} and the stellar-stream-based model of \citet{Ibata2024}. In addition, to examine how the stream disruption process depends on the assumed Galactic potential, we further considered an intermediate-mass model. The main characteristics of these three potentials are described below.

The first Milky Way mass model is taken from \citet{Jiao2021}, who fitted the \citet{Eilers2019} Galactic rotation curve based on \textit{Gaia} DR2. In this model, the Galaxy consists of a Plummer bulge, thin and thick stellar disks described by Miyamoto--Nagai profiles, and a dark matter halo represented by an Einasto profile. The halo density is given by
\begin{equation}
\rho_{\rm h}(r)=\rho_0 \exp\left[-\left(\frac{r}{h}\right)^{1/n}\right],
\end{equation}
where $\rho_0$ is the characteristic density, $h$ is the scale radius, and $n$ controls the shape of the radial decline. For their best-fit model, \citet{Jiao2021} adopted $n=5/3$, $\rho_0=3.473\times10^8\,M_\odot\,{\rm kpc}^{-3}$, and $h=0.3793~{\rm kpc}$, which gives a total Milky Way mass of $2.8\times10^{11}\,M_\odot$. We used this low-mass potential as Model~I, as it represents the rotation-curve-based mass scale and is consistent with the more recent \textit{Gaia} DR3 rotation-curve analysis of \citet{Jiao2023}.

As a comparison, we also adopted the mass model of \citet{Ibata2024}, which was constrained by fitting stellar streams. Their Galactic model contains six components: a fixed bulge, two fixed gas disks, thin and thick stellar disks described by double-exponential profiles, and a flattened dark halo described by a double-power-law density profile. The fixed bulge and gas disks were adopted from \citet{McMillan2017}. The halo density is given by
\begin{equation}
\rho_{\rm s}(s)=\rho_{0}
\left(\frac{s}{r_{0}}\right)^{-\gamma}
\left(1+\frac{s}{r_{0}}\right)^{\gamma-\beta}
\exp\left(-\frac{s^2}{r_{\rm t}^2}\right),
\end{equation}
with
\begin{equation}
s^2 = R^2+\frac{z^2}{q_{\rm m}^2},
\end{equation}
where $\rho_{0}$ is the characteristic density, $r_{0}$ is the scale radius, 
$r_{\rm t}$ is the truncation radius, $q_{\rm m}$ is the density flattening, 
and $\gamma$ and $\beta$ are the inner and outer logarithmic slopes, respectively.
For Model~II, we adopted the highest-likelihood \citet{Ibata2024} model with the outer halo slope fixed to $\beta=3$, as the alternative fit with $\beta$ free is strongly affected by the imposed upper limit on the virial mass. 
The adopted halo parameters are $\gamma=1.03$, $r_0=21.4~{\rm kpc}$, $q_{\rm m}=0.735$, and $M_{200}=1.24\times10^{12}\,M_\odot$.

Figure~\ref{fig:vc} compares the circular velocity curves of the adopted models with the rotation-curve measurements of \citet{Eilers2019}. The best-fit model of \citet{Jiao2021}, which was fitted to the \citet{Eilers2019} rotation curve, follows the data most closely. At large Galactocentric radii, $R>15~{\rm kpc}$, this best-fit rotation-curve model lies well below the stellar-stream best-fit model of \citet{Ibata2024}. 

To further explore globular-cluster disruption under the analytic-potential approximation, we also included a simplified intermediate-mass model, hereafter Model~III. This model consists of a Dehnen dark halo with $\gamma=0$, a Miyamoto--Nagai disk, and a Hernquist bulge. The dark halo has a mass of $M_{\rm h}=4.8\times10^{11}\,M_\odot$ and a scale radius of $a_{\rm h}=9.1~{\rm kpc}$. As shown in Fig.~\ref{fig:vc}, the outer-halo circular velocity of Model~III lies between those of the two models. Model~III is not designed to reproduce the rotation curve, but is included as a simplified reference model for testing tidal-shock effects. The baryonic-component parameters of the three models are summarized in Table~\ref{tab:mw_potentials}.

\subsection{Stream sample}

As shown in Fig.~\ref{fig:vc}, the discrepancy between rotation-curve- and stream-based Milky Way mass estimates becomes significant only beyond Galactocentric radii of $\sim 15$~kpc. In the stream-based analysis of \citet{Ibata2024}, the Galactic potential was constrained using 29 stellar streams, but only a small subset of these extend sufficiently far into the halo while also having enough identified members to support dynamical modeling. We therefore focused on the Pal~5 stream and the AAU stream, which are the best-constrained distant thin streams and are therefore well suited for testing outer Galactic potentials.

A key requirement for stream modeling is reliable six-dimensional phase-space information. While \textit{Gaia} provides highly precise sky positions and proper motions, distance measurements for distant and faint stream stars remain much more uncertain. RR~Lyrae stars are therefore especially valuable, as they are well-established standard candles and can provide distances with uncertainties of only a few percent. We used RRL-based distance information to improve the six-dimensional constraints of the modeled streams.

Pal~5 is a well-known disrupting globular cluster with tidal tails extending over $\sim20^\circ$ on the sky. We used the compiled Pal~5 stream-member catalog from \citet{Ibata2024} and \citet{Kuzma2022}, including the sky positions, proper motions, and line-of-sight velocities used in our comparison. The present-day cluster properties were taken from \citet{Vasiliev2021}. The distance constraint was guided by the RRL analysis of \citet{Price-Whelan2019}, who identified 27 RRL stars associated with the Pal~5 system, including 17 stream members, and measured a cluster distance of $20.6 \pm 0.2~{\rm kpc}$. 

For the second target, AAU, we used the corresponding stream-member compilation from \citet{Ibata2024}, supplemented by the spectroscopic constraints of \citet{Li2021}. We further used \textit{Gaia} DR3 RRL stars \citep{Clementini2023} along the stream to reconstruct its distance profile. Distances were estimated using the \textit{Gaia} $G_{\rm RP}$ period--luminosity--metallicity relation from \citet{Prudil2024},
\begin{equation}
    M_{G_{\rm RP}} = -1.464 \log_{10}(P) + 0.167[\mathrm{Fe/H}] + 0.113,
\end{equation}
which has the smallest intrinsic scatter among the three \textit{Gaia} bands considered by their calibration.

For the analytic-potential test, we additionally considered stream~\#17 from the catalog of \citet{Ibata2024}. Although this stream has relatively limited observational information and was not included in their Milky Way mass fitting, it is still of interest because its orbit is close to that of Sagittarius and can reach Galactocentric distances of up to $\sim80\ {\rm kpc}$. We therefore used a Sagittarius-like orbit as an approximate proxy, which allowed us to compare stream disruption on a distant outer-halo orbit with that on the closer Pal~5-like orbit.

\subsection{Stream modeling setup}

\begin{table*}
\centering
\setlength{\tabcolsep}{4pt}
\caption{
$N$-body simulations performed in this work.
}

\begin{tabular}{lllccccccc}
\hline
Run & System/orbit & MW model
& $M_{\rm prog}$ [$M_\odot$] & $r_{\rm h}$ [pc] 
& $m_{\rm GC}$ [$M_\odot$] & $m_{\rm MW}$ [$M_\odot$] 
& $\epsilon_{\rm GC}$ [pc] & $\epsilon_{\rm MW}$ [kpc]
& Duration \\
\hline

A1 & Sgr-like 
& Model~III 
& $10^4$ & 40 
& 1 & -- & 0.5 & -- & $12$ Gyr \\

A2 & Sgr-like 
& Model~III 
& $10^4$ & 40 
& 1 & $5.5\times10^3$ & 0.5 & 0.5 & $12$ Gyr \\

A3 & Pal~5-like 
& Model~III 
& $10^4$ & 27 
& 1 & -- & 0.5 & -- & $12$ Gyr \\

A4 & Pal~5-like 
& Model~III 
& $10^4$ & 27 
& 1 & $5.5\times10^3$ & 0.5 & 0.5 & $12$ Gyr \\

B1 & Pal~5 
& Model~I 
& $1.2\times10^5$ & 36 
& 1 & -- & 0.5 & -- & 2 Gyr \\

B2 & Pal~5 
& Model~II 
& $1.2\times10^5$ & 36 
& 1 & -- & 0.5 & -- & 2 Gyr \\

B3 & AAU 
& Model~I 
& $5\times10^3$ & 30 
& 1 & -- & 0.5 & -- & 3 Gyr \\

B4 & AAU 
& Model~II 
& $5\times10^3$ & 30 
& 1 & -- & 0.5 & -- & 5 Gyr \\

\hline
\end{tabular}
\tablefoot{
$M_{\rm prog}$ and $r_{\rm h}$ denote the initial mass and
initial half-mass radius of the progenitor globular cluster.
$m_{\rm GC}$ and $\epsilon_{\rm GC}$ denote the globular-cluster
particle mass and gravitational softening length, respectively.
For the live-Milky-Way simulations, $m_{\rm MW}$ and $\epsilon_{\rm MW}$
denote the Milky Way particle mass and gravitational softening length,
respectively.
}
\label{tab:simulations}
\end{table*}
 
We carried out gravity-only $N$-body simulations with the
\texttt{GIZMO} code \citep{Hopkins2015}. Orbit integrations and the
conversion between the adopted phase-space coordinates and initial
simulation coordinates were performed using \texttt{galpy}
\citep{Bovy2015} and \texttt{Agama} \citep{Vasiliev2019}. For the
observational comparison, we transformed the simulated and observed
stars into stream-aligned coordinates, using the Pal~5 coordinate frame
from \texttt{gala} \citep{Price-Whelan2017} and the AAU coordinate frame
from \texttt{galstreams} \citep{Mateu2023}.

Table~\ref{tab:simulations} summarizes the numerical simulations
performed in this work. The simulations are divided into two groups.
Runs A1--A4 are analytic-potential tests, designed to check
whether tidal shocks and the response of a live Milky Way model
produce stream features that are missed in a fixed analytic potential adopted by \citet{Ibata2024}.
Runs B1--B4 are stream-modeling simulations, in which Pal~5
and AAU are evolved in the two adopted Milky Way potentials and
compared directly with the available observations.

For the analytic-potential tests, we initialized a diffuse,
low-mass globular cluster on two representative orbits: a Pal~5-like
orbit and a Sagittarius-like orbit in the intermediate-mass Milky Way
model III. For each orbit, the same globular-cluster model and initial
orbital conditions were evolved in both a live $N$-body Milky Way
model and a fixed analytic Milky Way potential. For the live $N$-body runs, we adopted a high-resolution setup for which numerical convergence was verified using Milky Way-to-cluster particle mass ratios of 5500, 550, and 55, as well as gravitational softening lengths of 0.5 and 0.1 kpc for the Milky Way particles. The resulting stream morphology and dynamical trends remained stable across these numerical choices. These simulations were run for 12~Gyr to ensure complete disruption of the progenitor and to follow the subsequent evolution of the unbound debris.

For the stream-modeling simulations, we modeled the observed
Pal~5 and AAU streams in Galactic potentials derived from the rotation
curve and from stellar streams. Since the progenitor cluster of Pal~5 is
still observed and its present-day phase-space coordinates are relatively
well constrained, we initialized the progenitor from its observed
six-dimensional properties:
$(\alpha,\delta)=(229.019^\circ,-0.121^\circ)$,
$d_\odot=20.6\,{\rm kpc}$,
$(\mu_{\alpha},\mu_\delta)=(-2.730,-2.654)\,{\rm mas\,yr^{-1}}$,
and $v_{\rm los}=-58.6\,{\rm km\,s^{-1}}$.

For AAU, no surviving progenitor is observed, so the present-day
phase-space coordinates of the progenitor cannot be fixed directly.
We first estimated a local phase-space point from a robust fit to the
observed stream track and kinematics, and then explored nearby initial
conditions around this reference point. For each Galactic potential, the adopted initial conditions were chosen to reduce the dominant residuals
and to provide a reasonable compromise among the sky position,
heliocentric distance, proper motions, and line-of-sight velocity.

For the \citet{Jiao2021} potential, we adopted
$(\alpha,\delta)=(20.185^\circ,-27.13^\circ)$,
$d_\odot=20.6\,{\rm kpc}$,
$(\mu_{\alpha},\mu_\delta)=(-0.043,-0.856)\,{\rm mas\,yr^{-1}}$,
and $v_{\rm los}=-106.9\,{\rm km\,s^{-1}}$.
For the \citet{Ibata2024} potential, we adopted the same sky position and
distance but used
$(\mu_{\alpha},\mu_\delta)=(-0.057,-0.876)\,{\rm mas\,yr^{-1}}$
and $v_{\rm los}=-108.53\,{\rm km\,s^{-1}}$.

The initial masses and half-mass radii of the two progenitors were
chosen to approximately reproduce the observed stream length and width,
while leaving a Pal~5 remnant consistent with the present-day
cluster. The simulation durations were chosen to cover the observed
extent of the streams in each adopted potential.

\section{Testing the analytic-potential approximation}

We first used Runs A1--A4 to examine whether the fixed analytic-potential approximation can bias stream-based mass constraints. This provided a useful consistency check for the stream-based mass model of \citet{Ibata2024}, where the Milky Way potential is described analytically.

A fixed analytic potential provides only the mean gravitational field of the Galaxy. It does not respond dynamically to the orbiting cluster or to the stripped debris, and therefore cannot capture the self-consistent exchange of energy and angular momentum between the cluster and the Milky Way components. In contrast, a live $N$-body halo allows the background potential to fluctuate and respond to the passage of the cluster and its debris. If these differences substantially affect cluster disruption or stream structure, they could contribute to the discrepancy between rotation-curve- and stream-based Milky Way mass estimates. We therefore compared globular-cluster disruption in fixed analytic and live $N$-body Milky Way models for two orbits: the Sagittarius-like (Runs A1 and A2) and Pal~5-like (Runs A3 and A4) orbits. The resulting streams were examined in physical and dynamical space, with the latter characterized by the $\Delta E$--eccentricity distribution, using representative snapshots at different orbital phases.

\subsection{Spatial comparison}

Globular clusters gradually lose stars under the Galactic tidal field, while gravitational shocks during pericentric passages can significantly enhance their disruption \citep{Aguilar1988}. Once unbound, stars drift away from the progenitor on slightly different orbits, progressively forming the leading and trailing tidal tails. The simulations were evolved for 12~Gyr; here we show representative snapshots selected near matched pericentric and apocentric phases after substantial tidal stripping has occurred. These two phases provide characteristic views of the stream during strong tidal forcing and the subsequent redistribution of the stripped material.

\begin{figure}[ht!]
\centering

\includegraphics[width=\columnwidth, trim=10 25 0 0, clip]{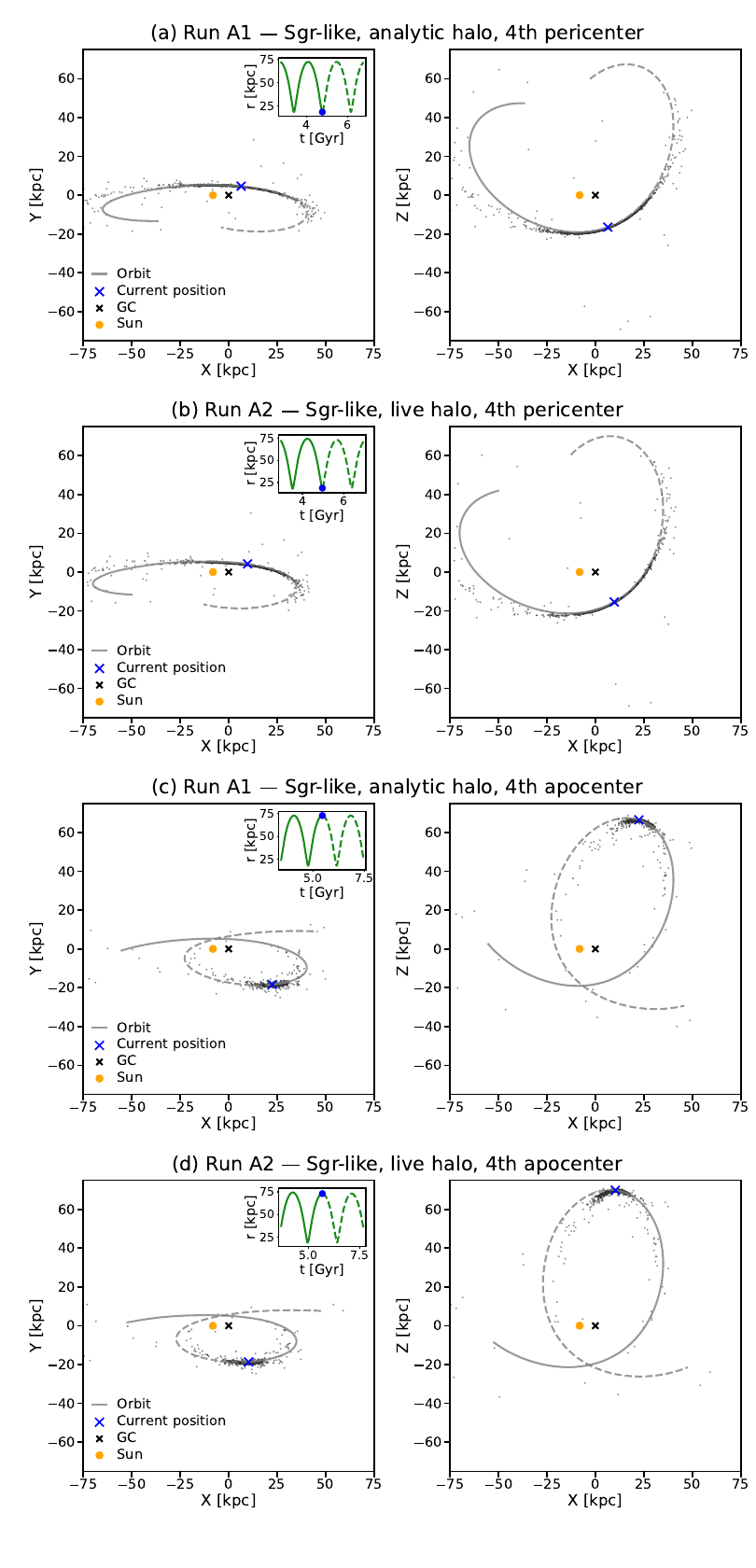}

\caption{Spatial comparison of the Sagittarius-like stream in the analytic and live $N$-body Milky Way potentials (Runs A1 and A2), shown at the fourth pericentric (\textit{a and b}) and apocentric (\textit{c and d}) passages. The gray curve shows the reference test-particle orbit; solid and dashed segments indicate the past and future trajectory, respectively. The blue cross marks the current position, and the inset gives the Galactocentric distance as a function of time. The Galactic disk lies in the $X$--$Y$ plane ($Z=0$), with the Galactic center at the origin and the Sun on the negative $X$-axis.}

\label{fig:sgr_spatial}
\end{figure}

For the Sagittarius-like orbit, we compare the stream evolution in the analytic and live $N$-body potentials in Fig.~\ref{fig:sgr_spatial}. At the fourth pericentric and apocentric passages, the two models produce very similar large-scale stream morphologies. The slight offset between the cluster trajectories indicates that weak fluctuations in the live $N$-body potential can produce small cumulative changes in the orbit over long timescales. Nevertheless, these orbital differences remain small and do not substantially alter the overall stream structure.

For the Pal~5-like orbit, the evolution is initially similar to that of the Sagittarius-like case. However, the Pal~5-like orbit has both a substantially shorter orbital period (approximately $0.3$~Gyr, compared with $\sim1.2$~Gyr for the Sagittarius-like orbit) and a smaller pericentric distance ($r_{\rm peri}\simeq9$~kpc compared with $\simeq18$~kpc). Consequently, over the same integration time it experiences both more frequent pericentric passages and stronger tidal shocks.

\begin{figure}[ht!]
\centering

\includegraphics[width=\columnwidth, trim=10 25 0 0, clip]{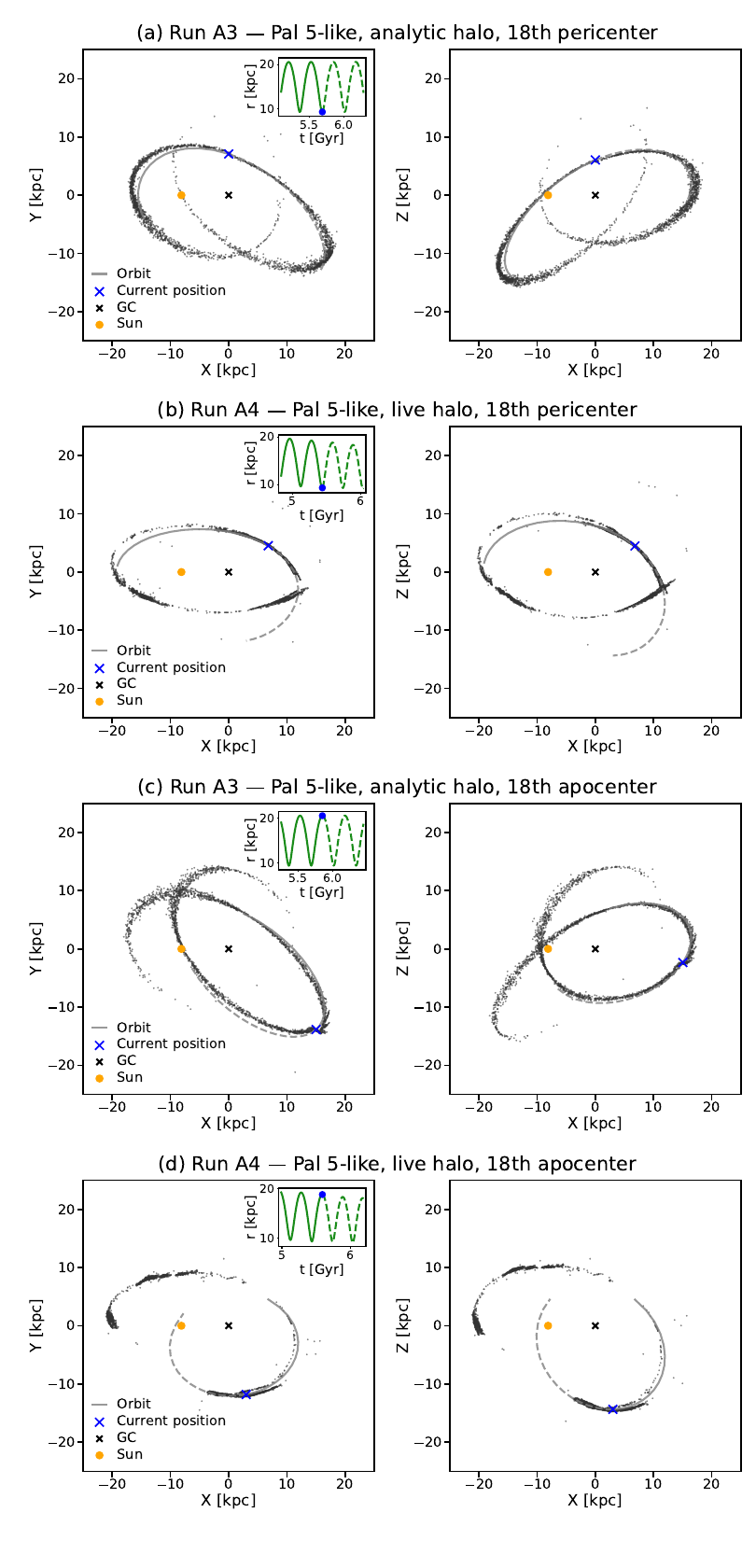}

\caption{Spatial comparison of the Pal~5-like stream in the analytic and live $N$-body Milky Way potentials (Runs A3 and A4), shown at the 18th pericentric (\textit{a and b}) and apocentric (\textit{c and d}) passages. The symbols and orbital curves are defined as in Fig.~\ref{fig:sgr_spatial}.}
\label{fig:pal5_spatial}
\end{figure}

As illustrated by the snapshots near the 18th pericenter and apocenter in Fig.~\ref{fig:pal5_spatial}, the differences between the analytic and live-halo runs are substantially larger than for the Sagittarius-like orbit. In the analytic-potential run, the stripped stars form a long, continuous, and relatively smooth stream. In the live-halo run, the debris becomes more spatially irregular, with visible bends, discontinuities, and partial overlap between material stripped during different passages.

Because individual snapshots provide only a phase-dependent view of the stream, we quantified the spatial difference between the two simulations throughout their evolution. The analytic and live-halo simulations were initialized with the same globular-cluster particle distribution and the same number of cluster particles. To quantify their spatial differences, we constructed two-dimensional particle-count maps in the $X$--$Y$ and $X$--$Z$ planes using the same fixed spatial range and a common $100\times100$ grid. Each map was smoothed with a Gaussian kernel with a standard deviation of one grid cell. For simulation $m\in\{\mathrm{ana},\mathrm{live}\}$ and projection $q\in\{XY,XZ\}$, we denote the Gaussian-smoothed particle count in grid cell $(i,j)$ by $N^{m,q}_{ij}$ and define the corresponding normalized spatial distribution as

\begin{equation}
P^{m,q}_{ij}
=
\frac{N^{m,q}_{ij}}
{\displaystyle\sum_{i,j}N^{m,q}_{ij}}.
\end{equation}

\begin{figure}[ht!]
    \centering

    \begin{subfigure}{\columnwidth}
        \centering
        \includegraphics[width=\linewidth]{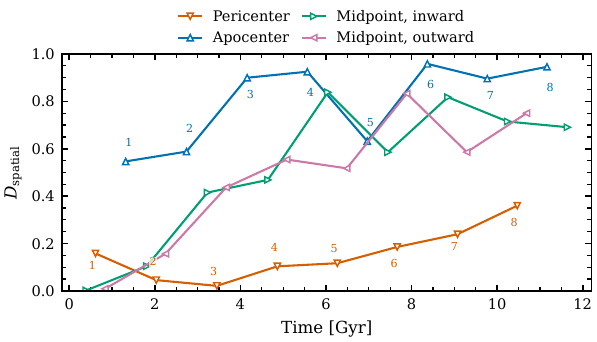}
        \caption{}
        \label{fig:d_spatial_sgr}
    \end{subfigure}

    \vspace{0.8em}

    \begin{subfigure}{\columnwidth}
        \centering
        \includegraphics[width=\linewidth]{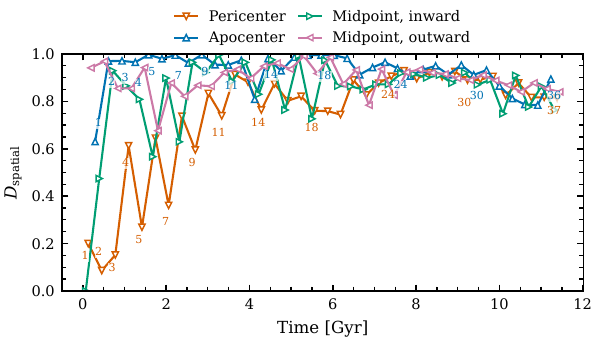}
        \caption{}
        \label{fig:d_spatial_pal5}
    \end{subfigure}

    \caption{
    Evolution of the spatial-distribution difference
($D_{\mathrm{spatial}}$) between the analytic and live-halo
    simulations. Values are measured at each pericentric and
    apocentric passage and at the inward and outward crossings of the
    midpoint radius
    $r_{\mathrm{mid}}=(r_{\mathrm{peri}}+r_{\mathrm{apo}})/2$.
    \textit{Panel (a)}: Sagittarius-like orbit. \textit{Panel (b)}:
    Pal~5-like orbit. Numbers indicate the successive pericentric and
    apocentric passages.
    }
    \label{fig:d_spatial_evolution}
\end{figure}

We then defined the spatial difference in each projection as

\begin{equation}
D_q
=
\frac{1}{2}
\sum_{i,j}
\frac{
\left(
P^{\mathrm{ana},q}_{ij}
-
P^{\mathrm{live},q}_{ij}
\right)^2
}{
P^{\mathrm{ana},q}_{ij}
+
P^{\mathrm{live},q}_{ij}
}
\end{equation}

\noindent and combined the two projected distances as

\begin{equation}
D_{\mathrm{spatial}}
=
\frac{1}{2}
\left(D_{XY}+D_{XZ}\right).
\end{equation}
By construction, $0\leq D_{\mathrm{spatial}}\leq1$,\footnote{For two normalized non-negative distributions, $(P_1-P_2)^2/(P_1+P_2)\leq P_1+P_2$. Summing over all grid cells therefore gives a maximum value of 2, which is reduced to unity by the prefactor $1/2$.} with smaller values indicating more similar spatial distributions and larger values indicating stronger spatial differences. We calculated $D_{\mathrm{spatial}}$ at each pericentric and apocentric passage within the simulated time range, as well as at two intermediate orbital phases corresponding to the inward and outward crossings of the midpoint radius, $r_{\mathrm{mid}}=(r_{\mathrm{peri}}+r_{\mathrm{apo}})/2$. The resulting evolution of $D_{\mathrm{spatial}}$ is shown in Fig.~\ref{fig:d_spatial_evolution}.

For the Sagittarius-like orbit, $D_{\mathrm{spatial}}$ is generally small near pericenter, indicating that the analytic and live-halo particle distributions remain most similar at this orbital phase. The values are substantially larger near apocenter, even at the fourth apocenter shown in Fig.~\ref{fig:sgr_spatial}, where the large-scale stream morphologies appear visually similar. At intermediate orbital phases, $D_{\mathrm{spatial}}$ generally lies between the pericentric and apocentric values. Because $D_{\mathrm{spatial}}$ is sensitive to spatial offsets between the two particle distributions, including those caused by slightly different orbital trajectories, a large value does not necessarily imply a large morphological difference. In contrast, a small value robustly indicates a high degree of spatial overlap. This sensitivity also contributes to the passage-to-passage fluctuations in $D_{\mathrm{spatial}}$. Nevertheless, comparing the same orbital phase across successive passages provides a useful measure of how the spatial overlap evolves with time. Despite some fluctuations, the overall level of $D_{\mathrm{spatial}}$ increases, indicating that differences between the analytic and live-halo simulations gradually accumulate during the evolution.

For the Pal~5-like orbit, $D_{\mathrm{spatial}}$ increases rapidly during the first $\sim2$~Gyr and subsequently remains large over most orbital phases. The repeated pericentric shocks make this inner orbit particularly sensitive to additional time-dependent fluctuations in the live $N$-body gravitational field, so that small perturbations accumulated over many orbital cycles can progressively alter the spatial distribution of the debris.

\subsection{Dynamical comparison}

To further assess the impact of the analytic-potential approximation, we compared the simulated streams in the $(E,L_z)$ plane. The distribution in the $(E,L_z)$ plane provides a useful diagnostic of how tidally stripped stars are distributed in energy and angular momentum, and of whether this distribution is modified when the Galactic potential is represented by a live halo rather than by a fixed analytic model. We further separated the leading and trailing arms to examine whether the two components exhibit distinct dynamical behavior.

\begin{figure}
    \centering
    \includegraphics[width=\columnwidth]{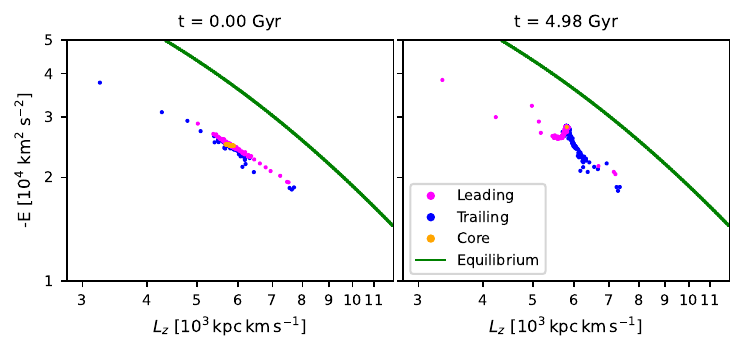}
    \caption{Energy--angular momentum distribution for the Sagittarius-like orbit (Run A1) at the initial stage ($t=0$) and at a later time ($t=4.98$~Gyr). The green curve shows the circular-orbit equilibrium sequence. Leading, trailing, and core particles are shown separately to illustrate how the dynamical distribution broadens as the system evolves.}
    \label{fig:ELz_2panel}
\end{figure}

Figure~\ref{fig:ELz_2panel} illustrates this behavior for the Sagittarius-like orbit by comparing the system at the initial stage and at a later time after complete tidal disruption. In both cases, the stars remain distributed along a relatively narrow sequence in the $(E,L_z)$ plane, although the spread around this sequence increases after disruption. The equilibrium curve gives the energy of a circular orbit at a given angular momentum and therefore provides a reference for the departure of individual stars from circular-orbit equilibrium. However, the detailed evolution is not particularly apparent in the $(E,L_z)$ plane, because the stars remain concentrated along a narrow locus. To characterize this departure more directly, we therefore considered the $\Delta E$--eccentricity plane, where $\Delta E$ is defined at each snapshot as the energy offset relative to the circular-orbit equilibrium energy at the same angular momentum. This quantity measures the excess energy above the circular-orbit value at fixed angular momentum and therefore correlates closely with orbital eccentricity.

\begin{figure}[ht!]
\centering
\includegraphics[width=\columnwidth, trim=15 10 25 10, clip]{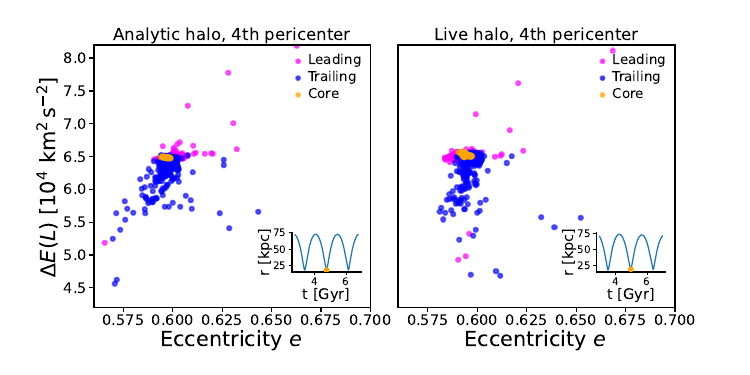}
\includegraphics[width=\columnwidth, trim=15 10 25 10, clip]{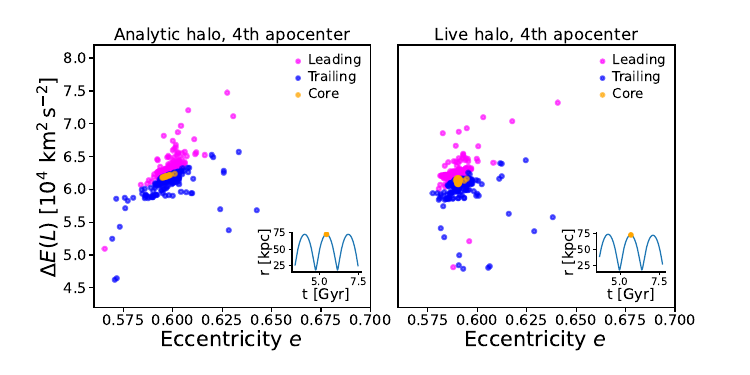}
\caption{
Distribution in the $\Delta E$--eccentricity plane for the Sagittarius-like stream. \textit{Top row}: Stream near pericenter. \textit{Bottom row}: Stream near apocenter. \textit{Left column}: Analytic-potential run, Run A1. \textit{Right column}: Live-halo run, Run A2.}
\label{fig:sgr_dEe}
\end{figure}

For the Sagittarius-like orbit (Fig.~\ref{fig:sgr_dEe}), although the spatial morphology is almost unchanged between the analytic and live-halo runs, a difference is still visible in the $\Delta E$--eccentricity plane. In the analytic-potential run, the distribution shows a clearer trend, with stars of lower eccentricity lying closer to the circular-orbit equilibrium sequence and therefore having smaller $\Delta E$. In the live-halo run, the distribution of stars is similar, but the trend is less clearly visible. The leading and trailing arms can still be distinguished, and they show similar distributions in the two cases.

\begin{figure}[ht!]
\centering
\includegraphics[width=\columnwidth, trim=15 10 25 10, clip]{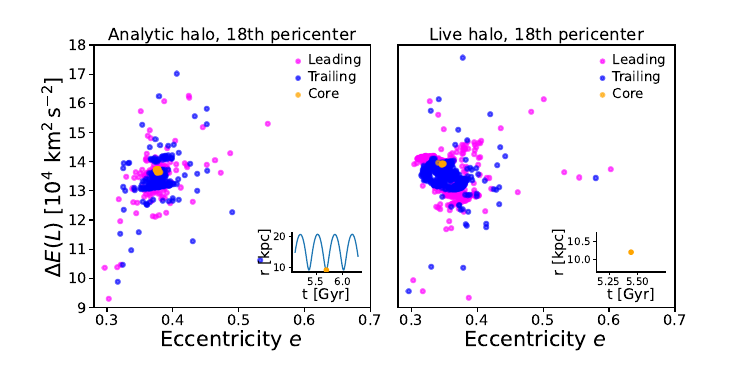}
\includegraphics[width=\columnwidth, trim=15 10 25 10, clip]{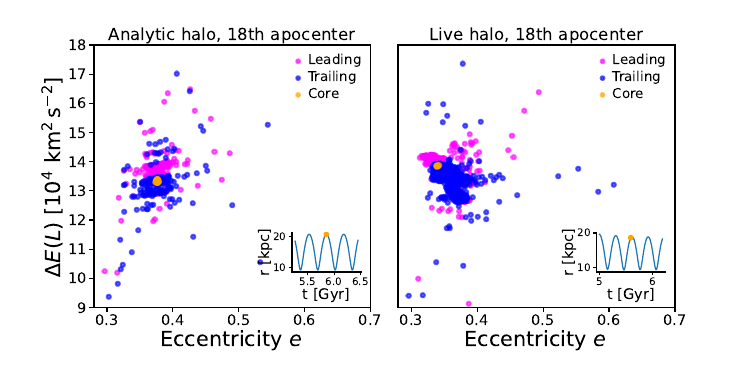}
\caption{
Distribution in the $\Delta E$--eccentricity plane for the Pal~5-like stream. \textit{Top row}: Stream near pericenter. \textit{Bottom row}: Stream near apocenter. \textit{Left column}: Analytic-potential run, Run A3. \textit{Right column}: Live-halo run, Run A4.
}
\label{fig:pal5_dEe}
\end{figure}

For the Pal~5-like orbit (Fig.~\ref{fig:pal5_dEe}), the difference between the two runs is larger. In the live-halo run, the distribution becomes more mixed and less organized than in the analytic-potential case, with the particles clustering into a poorly ordered structure and the separation between the leading and trailing arms becoming much less clear. 

\begin{figure}[ht!]
\centering

\begin{subfigure}{\columnwidth}
    \centering
    \includegraphics[width=\linewidth]{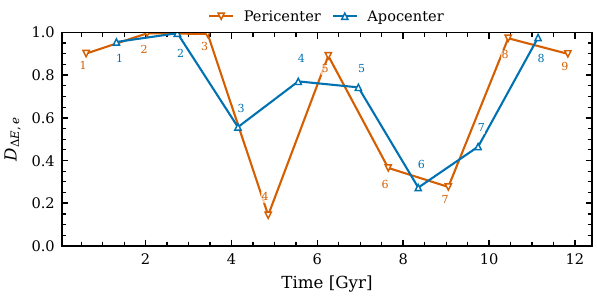}
    \caption{}
    \label{fig:d_dEe_sgr}
\end{subfigure}

\vspace{0.8em}

\begin{subfigure}{\columnwidth}
    \centering
    \includegraphics[width=\linewidth]{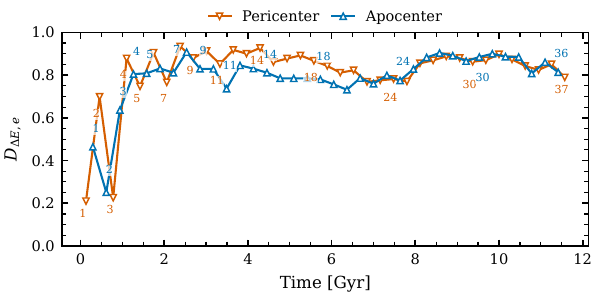}
    \caption{}
    \label{fig:d_dEe_pal5}
\end{subfigure}

\caption{
Evolution of the distribution difference
($D_{\Delta E,e}$) between the analytic and live-halo simulations
in the $\Delta E$--eccentricity plane.
Values are measured at successive pericentric and apocentric
passages. \textit{Panel (a)}: Sagittarius-like orbit. \textit{Panel (b)}: Pal~5-like orbit. Numbers indicate the successive pericentric
and apocentric passages.
}
\label{fig:d_dEe_evolution}

\end{figure}

To quantify these differences, we also calculated
$D_{\Delta E,e}$ from the particle distributions in the
$\Delta E$--eccentricity plane, using the same procedure as for the
spatial distributions. Its evolution at successive pericentric and
apocentric passages is shown in Fig.~\ref{fig:d_dEe_evolution}.
Unlike $D_{\mathrm{spatial}}$, no systematic difference is seen
between the pericentric and apocentric measurements, since
$\Delta E$ and eccentricity are less sensitive to the instantaneous
orbital phase than the spatial configuration of the stream. For the
Sagittarius-like orbit, $D_{\Delta E,e}$ varies considerably with
time, reaching both large and small values. For the
Pal~5-like orbit, $D_{\Delta E,e}$ also shows relatively large
fluctuations during the first $\sim1$~Gyr, but subsequently remains
generally high, indicating a more persistent difference between the
two simulations. This is consistent with the stronger differences
seen in the spatial and dynamical structure of the Pal~5-like
stream.

Taken together, these comparisons show that the differences between
the analytic and live-halo simulations are generally more pronounced
for the Pal~5-like orbit than for the Sagittarius-like orbit,
although differences in dynamical space remain visible in the latter.
For the two orbital configurations considered here, this is consistent
with the stronger tidal shocks expected for the smaller-pericenter
Pal~5-like orbit \citep{Aguilar1988,Hammer2024}, which accelerate
the disruption of the progenitor and generate an unbound stellar
population earlier and more efficiently. Once unbound, the debris
evolves more directly under the Galactic gravitational field and can
therefore become increasingly sensitive to differences between a
smooth analytic potential and a self-consistent live halo.

This tendency is opposite to that of the discrepancy between
rotation-curve- and stream-based Milky Way mass estimates in
Fig.~\ref{fig:vc}, which becomes larger toward the outer halo.
Therefore, although tidal shocks and the live response of the
Galactic halo can affect simulated stellar streams, they are unlikely
to be the dominant origin of the discrepancy between stream- and
rotation-curve-based mass estimates at $R_{\rm GC}>15~{\rm kpc}$.

\section{Distance constraints and stream modeling}
\subsection{Reevaluated distances}

Having tested the impact of the analytic-potential approximation, we next turned to a direct comparison between observed stellar streams and simulations in the two Milky Way potentials. The main question is whether the present-day phase-space properties of distant thin streams can distinguish between the low-mass rotation-curve-based potential of \citet{Jiao2021} and the high-mass stream-based potential of \citet{Ibata2024}. This provides a consistency test of the two mass models in the radial range where their circular-velocity curves begin to diverge.

For this comparison, accurate distance information is particularly important.
Pal~5 already has a carefully examined
RRL distance constraint from \citet{Price-Whelan2019}, whereas the
distance structure of AAU is less well established. We therefore first refined
the observational distance constraint for the AAU stream using RRL stars.
We identified candidate RRL members associated with the stream based on
their spatial and kinematic consistency with the stream track. Specifically,
we selected RRL stars that are aligned in stream coordinates
$(\phi_1,\phi_2)$ and have proper motions compatible with the AAU stream.

\begin{figure}
    \centering
    \includegraphics[width=\columnwidth]{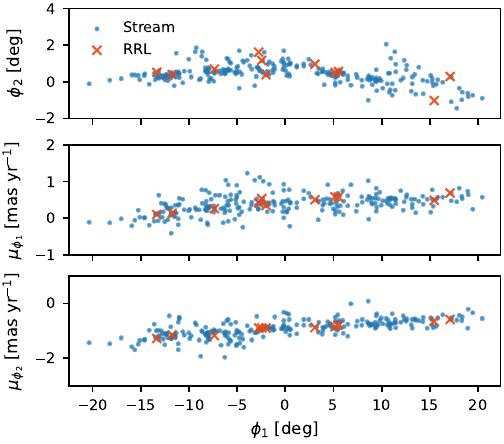}
    \caption{RRL candidates associated with the AAU stream, selected based on spatial and proper-motion consistency.}
    \label{fig:AAU_rrl}
\end{figure}

\begin{figure}
    \centering
    \includegraphics[width=\columnwidth]{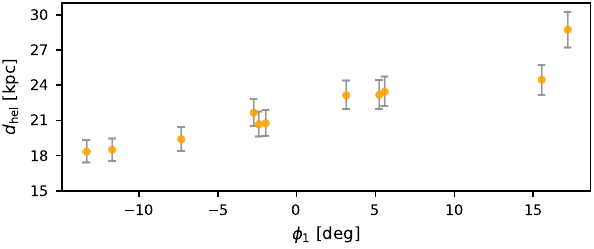}
    \caption{Distance profile of the AAU stream derived from RRL stars. Error bars indicate statistical uncertainties from the distance estimation.}
    \label{fig:AAU_dist}
\end{figure}

The resulting candidate sample is shown in Fig.~\ref{fig:AAU_rrl}. Although the number of candidates is limited, they approximately follow a coherent trend along the stream, suggesting that they are plausible RRL members of the AAU system. Using these stars, we derived distances along the stream from the period--metallicity--luminosity relation and estimated the corresponding uncertainties. The resulting distance profile is shown in Fig.~\ref{fig:AAU_dist}. The inferred uncertainties are typically $\sim 1$~kpc for individual stars, corresponding to a fractional distance uncertainty of a few percent.

Our selection is similar to that of \citet{Dominguez2026}, but differs in the treatment of the most distant candidate. While that study assigns this star a relatively high membership probability ($p_{\rm memb} > 0.5$), it excludes the star on the grounds that it lies offset from the stream track in celestial coordinates, whereas we retained it in our sample. Our inferred distance trend along the stream is nevertheless in good agreement with their results.

\begin{figure*}
    \centering
    \begin{subfigure}{0.49\textwidth}
        \centering
        \includegraphics[width=\linewidth]{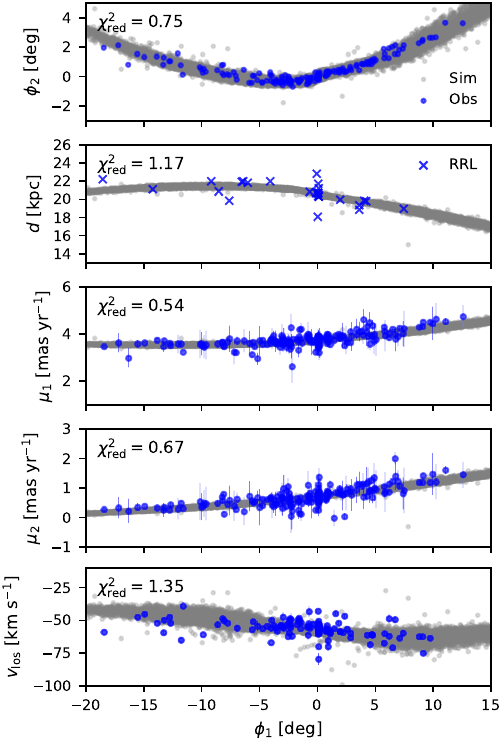}
        \caption{}
        \label{fig:pal5_mlow_chi2}
    \end{subfigure}
    \hfill
    \begin{subfigure}{0.49\textwidth}
        \centering
        \includegraphics[width=\linewidth]{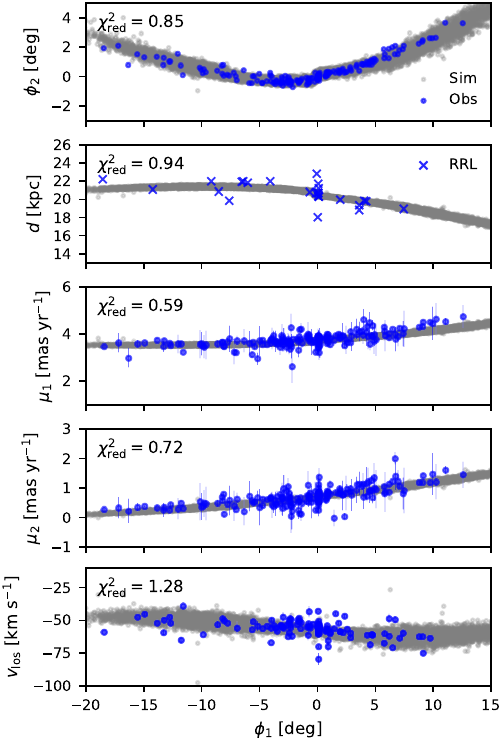}
        \caption{}
        \label{fig:pal5_mibata_chi2}
    \end{subfigure}

    \caption{Simulated Pal~5 stream compared with the observations in two Galactic potentials. From top to bottom, the panels show $\phi_2$, heliocentric distance, $\mu_1$, $\mu_2$, and $v_{\rm los}$ as functions of $\phi_1$ in stream coordinates for the \citet{Jiao2021} potential, Run B1 (\textit{Col. a}) and the \citet{Ibata2024} potential, Run B2 (\textit{Col. b}). Gray points denote simulated stream particles, blue points with error bars denote observations, and blue crosses mark the RRL distance constraints.}

    \label{fig:pal5_5panel_chi2}
\end{figure*}
\subsection{Modeling streams in different potentials}

We next used the distance constraints derived above, together with the
available astrometric and spectroscopic measurements, to compare the
simulated Pal~5 and AAU streams with the observations in Runs B1--B4.

\subsubsection{Palomar 5}

Figure~\ref{fig:pal5_5panel_chi2} compares the simulated Pal~5 stream
with the observations in two Galactic potentials. The left panel shows
the result obtained in the low-mass rotation-curve-based
potential of \citet{Jiao2021} (Run B1), whereas the right panel shows
the corresponding result in the high-mass stream-based potential of
\citet{Ibata2024} (Run B2).
In both cases, the simulated stream is compared with the observed
Pal~5 members, the RRL distance measurements of
\citet{Price-Whelan2019}, and the line-of-sight velocity measurements
from \citet{Ibata2024} and \citet{Kuzma2022}, after transforming both
the model and the data into the Pal~5 stream-coordinate system. The
RRL points are color-coded by the membership probabilities
reported by \citet{Price-Whelan2019}. The
comparison is shown in five projected dimensions: stream latitude
($\phi_2$), heliocentric distance ($d$), the two proper-motion components
$\mu_1$ and $\mu_2$, and line-of-sight velocity ($v_{\rm los}$), all as
functions of stream longitude ($\phi_1$).

Overall, both rotation-curve-based and stream-based potentials 
reproduce the observed morphology and the main
proper-motion trends of the Pal~5 stream well. The
heliocentric-distance panel shows a similar large-scale trend in the
simulations and in the RRL measurements, although several
RRL stars lie outside the main simulated stream distribution. This
may partly reflect the limited number of RRL tracers, together with
residual uncertainties in their membership probabilities and distance
estimates. In the line-of-sight velocity plane, some observed stars are
also not fully covered by the simulated streams in either potential. This
may be partly due to the scatter and possible systematics in the
available spectroscopic measurements, which are compiled from
heterogeneous datasets.

To quantify how well the two potentials reproduce the observed Pal~5
stream, we computed $\chi^2$-like statistics in the five observational
planes shown in Fig.~\ref{fig:pal5_5panel_chi2}. For each plane, each
observed data point is compared with the local mean of the simulated
stream at the corresponding $\phi_1$. We defined
\begin{equation}
\chi^2_{\rm red} =
\frac{1}{N_y-1}
\sum_{i=1}^{N_y}
\frac{
\left(y_{{\rm obs},i}-y_{{\rm sim},i}\right)^2
}{
\sigma_{{\rm obs},i}^2+s_{{\rm obs},i}^2+s_{{\rm sim},i}^2
},
\end{equation}
where $y$ denotes $\phi_2$, $d_\odot$, $\mu_1$, $\mu_2$, or
$v_{\rm los}$, $N_y$ is the number of observed data points in that
plane, $\sigma_{{\rm obs},i}$ is the measurement uncertainty, and
$s_{{\rm obs},i}$ and $s_{{\rm sim},i}$ describe the local scatter of
the observed and simulated stream, respectively. The resulting values
are shown in the upper left of each panel.

The reduced $\chi^2$ values indicate only small differences between the
two models. In the sky-track plane, where the \textit{Gaia} positions provide
the strongest constraints, the \citet{Jiao2021} potential gives a smaller
residual, with $\chi^2_{\rm red}(\phi_2)=0.75$, compared to
$\chi^2_{\rm red}(\phi_2)=0.85$ for the \citet{Ibata2024} potential. The
\citet{Jiao2021} potential also gives slightly smaller residuals in the
proper-motion planes, whereas the \citet{Ibata2024} potential gives
smaller residuals in heliocentric distance and line-of-sight
velocity. However, the heliocentric-distance and line-of-sight velocity planes
show larger residuals in both potentials. The deviations occur in
similar regions of $\phi_1$ for the two models, suggesting that they are
not mainly driven by the choice of Galactic potential but instead
reflect limitations in the available distance and spectroscopic constraints.
Taken together, the preferences are not uniform across the five
observational planes. The current Pal~5 data therefore do not strongly
discriminate between the two Galactic potentials over the radial range
probed by the stream, $R_{\rm GC}\sim15$--$17~{\rm kpc}$.

\subsubsection{ATLAS--Aliqa Uma}

\begin{figure*}
    \centering
    \begin{subfigure}{0.49\textwidth}
        \centering
        \includegraphics[width=\linewidth]{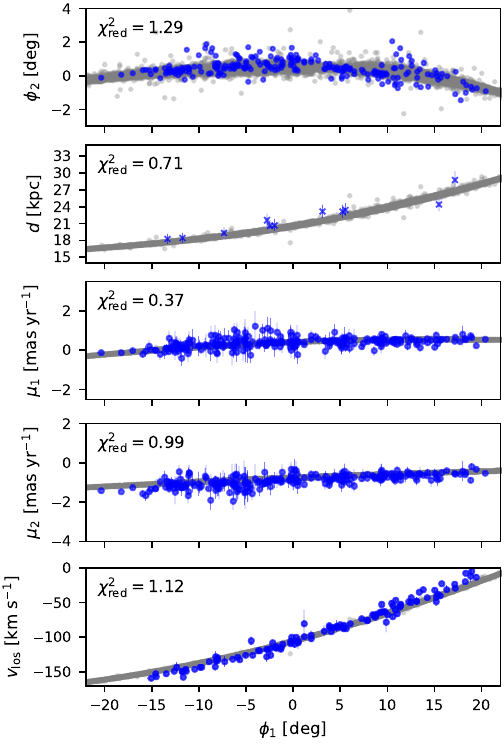}
        \caption{}
        \label{fig:AAU_model_low_chi2}
    \end{subfigure}
    \hfill
    \begin{subfigure}{0.49\textwidth}
        \centering
        \includegraphics[width=\linewidth]{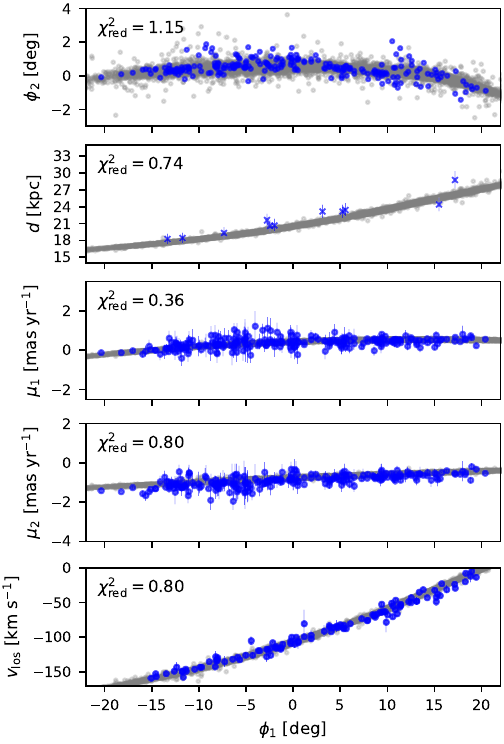}
        \caption{}
        \label{fig:AAU_model_ibata_chi2}
    \end{subfigure}

    \caption{Same as Fig. \ref{fig:pal5_5panel_chi2} but for the simulated AAU stream and with the \citet{Jiao2021} potential, Run B3 (\textit{Col. a}), and the \citet{Ibata2024} potential, Run B4 (\textit{Col. b}). }
    \label{fig:AAU_model}
\end{figure*}

The AAU stream provides a complementary test to
Pal~5: it extends over a larger angular range and reaches Galactocentric
distances of nearly 30~kpc, although its observational constraints are
more limited. Figure~\ref{fig:AAU_model} compares the simulated AAU
stream with the observations in two Galactic potentials. The left panel
shows the result obtained in the best-fit, low-mass
rotation-curve-based potential of \citet{Jiao2021} (Run B3), while the
right panel shows the corresponding result in the stream-based potential
of \citet{Ibata2024} (Run B4). The comparison is shown in the
stream-coordinate system, including $\phi_2$, heliocentric distance,
the two proper-motion components, and line-of-sight velocity as
functions of $\phi_1$. The distance constraints are derived from the
RRL stars discussed in Sect.~4.1, and the line-of-sight velocity
measurements are taken from the compiled catalog of \citet{Ibata2024}
and from \citet{Li2021}.

As discussed in Sect.~2, AAU has no surviving progenitor, so its
initial conditions have to be inferred from the observed stream itself.
For each Galactic potential, we therefore explored nearby initial
conditions around the phase-space point inferred from the observed
stream, and adopted the set that gives the best compromise among the
different observational projections. The details of this initial-condition
exploration are given in Appendix~\ref{app:aau}.
Moreover, AAU is less regular than Pal~5. \citet{Li2021} showed that
ATLAS and Aliqa Uma are very likely two segments of a single stream,
but with discontinuities in both morphology and density, including the
kink feature and broadening/gap structures. These make AAU
intrinsically difficult to reproduce with a single smooth-stream model
in a static analytic potential.

We used the result shown in Fig.~\ref{fig:AAU_model} primarily as a comparison
of the broad phase-space trends predicted by the two Galactic
potentials. In both potentials, the simulations
reproduce the main trends in sky position, heliocentric
distance, $\mu_1$, $\mu_2$, and line-of-sight velocity. Some residuals remain, especially in the sky-track and $\mu_2$
projections, and part of the simulated debris lies outside the observed
stream track in the $\phi_1$--$\phi_2$ plane. This mismatch may arise
from the perturbed nature of AAU, as well as from limitations of using a
single progenitor evolved in a static analytic potential. In such a
model, time-dependent tidal effects and external perturbations are not
fully captured, which may affect the predicted width, track, and
phase-space distribution of the stream debris.

We computed the same reduced $\chi^2$-like statistics for AAU as for
Pal~5, with the resulting values shown in the upper left of each panel
in Fig.~\ref{fig:AAU_model}. The two potentials give very similar
residuals in all five observational projections. The \citet{Ibata2024}
potential gives slightly smaller residuals in the sky-track and
line-of-sight velocity planes, with
$\chi^2_{\rm red}(\phi_2)=1.15$ and
$\chi^2_{\rm red}(v_{\rm los})=0.80$, compared to 1.29 and 1.12 for the
\citet{Jiao2021} potential. It also gives marginally smaller residuals
in the two proper-motion components. By contrast, the
\citet{Jiao2021} potential gives a slightly smaller distance residual. These differences are small, and no
projection shows a strong preference for either potential.

For AAU, the small differences between the two potentials are therefore
subdominant to the uncertainties associated with the progenitor initial
conditions and the perturbed stream morphology. Although the
\citet{Ibata2024} potential gives slightly smaller residuals in most
projections, this preference is not strong enough to identify it as a
clearly better model for AAU over $R_{\rm GC}\sim18$--$30\,{\rm kpc}$.
Together with the Pal~5 comparison, this suggests that both the
rotation-curve-based and stream-based Galactic potentials can be
broadly compatible with the observed properties of distant thin stellar
streams over $R_{\rm GC}\sim15$--$30~{\rm kpc}$. At the same time, the
limited number of well-constrained distant streams, together with the
remaining uncertainties in their distances, kinematics, and progenitor
properties, means that the present stream phase-space information does
not yet provide sufficiently strong constraints to distinguish decisively
between these Galactic potentials.

\section{Discussion and conclusion}

In this work, we first tested the use of static analytic potentials in
stream modeling. Previous studies have shown that time dependence and halo complexity can affect stream evolution and potential recovery \citep{Buist2015,Bonaca2014,Sanderson2017}. We extended these studies through a controlled high-resolution N-body comparison in which the same self-gravitating globular cluster was evolved in matched analytic and live Galactic models, allowing the impact of replacing a fixed analytic potential with a live $N$-body realization to be assessed under matched initial conditions. Our results show that the tension between rotation-curve- and stream-based Milky Way mass estimates is unlikely to be caused solely by this approximation. Live-halo effects are most evident for an inner, Pal~5-like orbit, where repeated pericentric passages produce stronger tidal shocks and more irregular debris. However, these effects are much weaker for a more distant, Sagittarius-like orbit. This trend is opposite to the radius range in which the mass-profile discrepancy becomes most pronounced, suggesting that the static-potential approximation alone cannot explain the difference between the two types of mass estimates.

We then examined whether present-day observations of thin stellar
streams can distinguish between rotation-curve-based and
stream-based Galactic potentials. Before performing this comparison,
we revisited the distance constraints for Pal~5 and
AAU using RRL stars in order to refine the
observational constraints. We find that both potentials tested here
reproduce the main observed trends of Pal~5 and AAU over $R_{\rm GC}\sim15$--$30~{\rm kpc}$. The current data
therefore do not strongly favor one potential over the other in this
radial range. Earlier studies of Pal~5 also illustrate the sensitivity of stream constraints to the adopted data and modeling assumptions. \citet{Kupper2015} obtained relatively tight Galactic mass constraints using density features along the stream, although some of these features were later questioned with deeper photometry \citep{Ibata2016,Thomas2016}. Our analysis addresses the more specific question of whether the current observational constraints can distinguish between the two Galactic potentials considered here. In the following, we discuss two main limitations that are important for interpreting this result: distance uncertainties and model-dependent extrapolations.

\subsection{Impact of distance uncertainties on the inferred Galactic potential}

For the streams considered here, the distance scale remains an important
source of uncertainty in constraining the Milky Way potential. Sky
positions are measured very precisely, and \textit{Gaia} provides strong
proper-motion constraints, whereas heliocentric distances remain less precisely constrained. Even for RRL stars, which are standard-candle tracers, the
typical relative distance precision is $\sim3\%$. For the most distant RRL candidates in the
AAU region, this corresponds to an uncertainty of up to
$\sim1$~kpc.

This scale is non-negligible because the distance determines the
Galactocentric radius at which the stream is assumed to probe the
potential. As a simple illustration, if one interprets the circular velocity in terms of the enclosed mass under a spherical approximation, one obtains
\begin{equation}
    v_{\rm c}^2(R) = \frac{G M(<R)}{R},
\end{equation}
or equivalently
\begin{equation}
    M(<R) = \frac{R v_{\rm c}^2}{G}.
\end{equation}
Although this expression is only illustrative and does not represent the
full stream-modeling problem, it shows that a change in the adopted
distance scale directly shifts the radius at which the mass profile is
sampled, and therefore can affect the inferred mass scale.

This issue is relevant even for Pal~5, which is among the best
constrained streams at $R_{\rm GC}>15\, {\rm kpc}$. Previous studies have
adopted different heliocentric distances for the Pal~5 cluster. The
commonly used Harris-catalog value \citep[2010 edition]{Harris1996} is $d\simeq23.2\, {\rm kpc}$, while
\citet{Kupper2015} obtained
$d=23.58^{+0.84}_{-0.72}\,{\rm kpc}$ from stream modeling.
In contrast, \citet{Price-Whelan2019} derived an RRL-based
distance of $d=20.6\pm0.2\,{\rm kpc}$, and
\citet{BaumgardtVasiliev2021} reported
$d=21.941^{+0.520}_{-0.508}\,{\rm kpc}$. In the stream-fitting analysis of
\citet{Ibata2024}, the distance of Pal~5 was allowed to vary within
the uncertainties of \citet{BaumgardtVasiliev2021}. This distance
range is larger than the difference between the heliocentric-distance
predictions of the two Galactic potentials shown in
Fig.~\ref{fig:pal5_5panel_chi2}. Thus, even for Pal~5, current distance
uncertainties can mask the relatively small differences between the
models. More precise and homogeneous distance measurements along
the stream would therefore be important when using streams to
discriminate between Galactic mass models.

\subsection{Radial ranges and profile-dependent extrapolations}

The comparison presented in this work is limited to the radial range sampled by Pal~5 and AAU, $R_{\rm GC}\sim15$--$30\,{\rm kpc}$. The constraints in this range should not be interpreted as a direct validation of either mass model at larger radii.

Disk rotation curves can be used to infer the radial force from the kinematics of disk tracers, usually through the Jeans equation under assumptions of axisymmetry and approximate dynamical equilibrium. They therefore only provide a relatively direct constraint on the circular velocity over the radial interval covered by the tracer population. Recent \textit{Gaia}-based Milky Way rotation curves reach the outer disk, typically $R_{\rm GC}\simeq25$--$30\,{\rm kpc}$ \citep{Eilers2019,Jiao2021,Jiao2023,Ou2024}. Any inference of the halo mass at substantially larger radii requires an extrapolation from the measured disk region to the virial scale. The result of this extrapolation depends on the adopted dark-matter profile. In particular, \citet{Jiao2021} showed that Navarro–Frenk–White (NFW) or generalized NFW profiles can introduce a methodological bias when fitting a declining Milky Way rotation curve, because different halo profiles can reproduce the measured rotation curve over $R_{\rm GC}\lesssim25$--$30\,{\rm kpc}$ while implying different masses at larger radii.

A similar caveat applies to the stream-based curve. The
\citet{Ibata2024} curve shown in Fig.~\ref{fig:vc} is the
circular-velocity profile of their best-fitting global mass model, rather
than a point-by-point rotation-curve measurement. The stream data
constrain the potential through the observed phase-space locations of
the streams and through the orbits allowed by the adopted model.
Therefore, the radial coverage of the constraint is set by the stream
sample and is not uniform in Galactocentric radius. In the fitted stream
sample, AAU is among the most distant streams and
reaches only $\sim30\,{\rm kpc}$. At radii beyond those directly sampled
by the fitted streams, the inferred circular-velocity curve remains
sensitive to the adopted halo parametrization and priors. This is
analogous to the profile dependence in rotation-curve extrapolations,
although the underlying tracers and fitting procedures are different.

In this work, the comparison with Pal~5 and AAU provides a local test of the low-mass rotation-curve-based potential and the higher-mass stream-based potential over $R_{\rm GC}\sim15$--$30\,{\rm kpc}$. At larger radii, neither rotation-curve extrapolations nor the present stream constraints provide a decisive determination of the Milky Way mass profile. Nevertheless, the comparison presented here shows how stellar streams
can be used not only to test existing Galactic potentials, but also to
refine them. Quantitative residuals based on stream morphology and
kinematics provide a natural basis for fitting improved mass models,
provided that the stream constraints are sufficiently precise and that
the origin of complex stream features is properly modeled. This is especially relevant for streams such as
AAU, where the observed morphology can reflect external perturbations or progenitor properties in addition to the smooth Milky Way potential. More distant tracers, such as the Sagittarius stream, might be important for anchoring the outer-halo mass profile, although their modeling is also affected by the Large Magellanic Cloud perturbation \citep{Vasiliev2021b} and by the gas content or past infall history of the Sagittarius progenitor \citep{Wang2022}.

\begin{acknowledgements}
We thank the anonymous referee for the constructive comments and suggestions that helped improve this work. Simulations in this work were performed using the high-performance computing (HPC) resources of MesoPSL, financed by the Equip@Meso project (reference ANR-10-EQPX-29-01) of the ``Investissements d'Avenir'' program supervised by the Agence Nationale de la Recherche.

This work was granted access to HPC computing and storage resources
by GENCI at IDRIS under the grant 2025-AD010416448 on the Jean Zay
supercomputer CSL partition.

M.C. acknowledges financial support from the China Scholarship Council
(CSC) for the PhD fellowship. M.C. is also grateful to the PLATO mission, Piercarlo Bonifacio, and Benoît Mosser for the additional support and assistance that made the start of this PhD project possible.
\end{acknowledgements}

\bibliographystyle{aa}
\bibliography{references}

\begin{appendix}

\nolinenumbers

\section{AAU stream with different progenitor initial conditions}
\label{app:aau}

\begin{figure*}
    \centering
    \begin{subfigure}{0.49\textwidth}
        \centering
        \includegraphics[width=\linewidth]{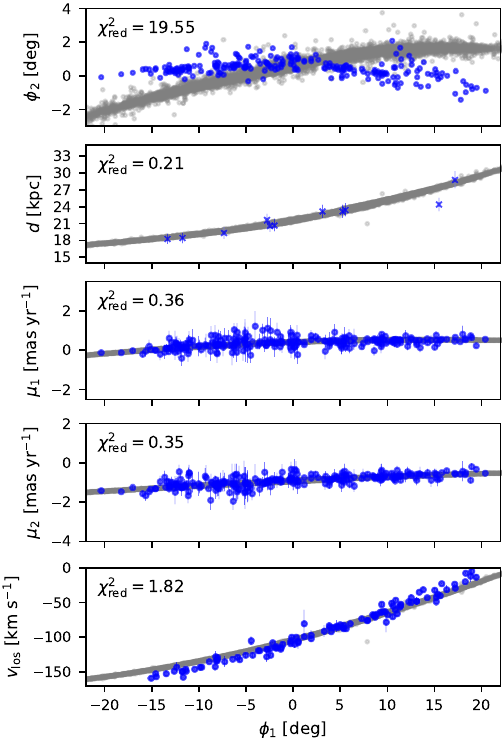}
        \caption{}
        \label{fig:aau_ic_direct_jiao}
    \end{subfigure}
    \hfill
    \begin{subfigure}{0.49\textwidth}
        \centering
        \includegraphics[width=\linewidth]{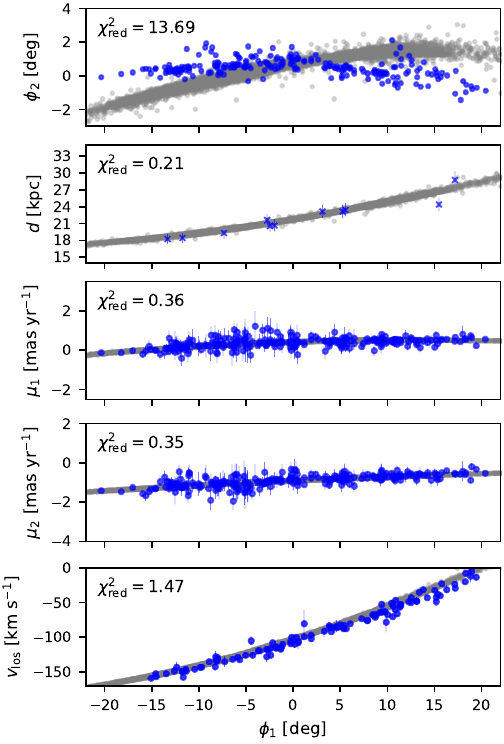}
        \caption{}
        \label{fig:aau_ic_direct_ibata}
    \end{subfigure}

    \caption{AAU stream models obtained in
    Test 1, corresponding to the phase-space point directly inferred from
    the local stream fit. The panels show, from top to bottom,
    $\phi_2$, heliocentric distance, $\mu_1$, $\mu_2$, and
    $v_{\rm los}$ as functions of $\phi_1$, for the \citet{Jiao2021} potential (\textit{Col. a}) and the \citet{Ibata2024} potential (\textit{Col. b}). Gray points are simulated stream particles, blue points with error bars denote observations, and blue crosses mark the RRL distance constraints.}
    \label{fig:aau_ic_direct_fit}
\end{figure*}

AAU has no surviving progenitor, and the present-day phase-space
coordinates of the disrupted system are therefore not directly fixed by
observations. We therefore explored several nearby choices of the
progenitor phase-space point, in order to identify a starting point for
the stream model that gives a reasonable match to the available
observables. These tests also provide a useful check of how sensitive
the AAU model is to the adopted progenitor phase-space point.

The present-day progenitor phase-space points and the corresponding
reduced $\chi^2$-like statistics are listed in
Table~\ref{tab:aau_ic_tests}. The coordinates and proper motions are
given in the AAU stream-coordinate system. Since $\phi_1$ is used as
the independent coordinate in the comparison, no $\chi^2_{\rm red}$ value is
reported for this quantity.

Test 1 uses the phase-space point obtained directly from the local fit to
the stream, and the resulting models are shown in
Fig.~\ref{fig:aau_ic_direct_fit}. This choice gives reasonable agreement
in distance, proper motion, and line-of-sight velocity, but produces a
large offset in the sky-track plane, with
$\chi^2_{\rm red}(\phi_2)=19.55$ for the \citet{Jiao2021} potential and 13.69
for the \citet{Ibata2024} potential. Since this offset is mainly related
to the projected orbital direction, we then varied $\mu_2$ and distance
separately. Changing only $\mu_2$ in Test 2 strongly improves the
sky-track residuals, but worsens the agreement in $\mu_2$ itself.
Changing only the distance in Test 3 also improves the sky track, but
requires a distance of $17.50~{\rm kpc}$ and becomes inconsistent with
the RRL distance constraints.

Tests 4 and 5 therefore adopt intermediate choices. In Test 4, both the
distance and $\mu_2$ are adjusted, giving much smaller sky-track
residuals while keeping the distance and proper-motion residuals at a
reasonable level. In Test 5, the same distance and proper motions are
kept, but the initial line-of-sight velocity is slightly adjusted to reduce
the systematic offset in the $v_{\rm los}$ plane. This final set of initial
conditions is used for the AAU models discussed in the main text.

\begin{table}
\centering
\caption{Present-day progenitor phase-space points and reduced
$\chi^2$-like statistics for the AAU stream models.}
\label{tab:aau_ic_tests}
\begin{tabular}{lcccc}
\hline
Quantity
& \multicolumn{2}{c}{Jiao+2021}
& \multicolumn{2}{c}{Ibata+2024} \\
\cline{2-5}
& IC & $\chi^2_{\rm red}$
& IC & $\chi^2_{\rm red}$ \\
\hline
\multicolumn{5}{l}{Test 1: direct local fit} \\
$\phi_1$ [deg]        & $-0.533$ & --    & $-0.533$ & --    \\
$\phi_2$ [deg]        & $0.649$  & 19.55 & $0.649$  & 13.69 \\
$d_\odot$ [kpc]       & $21.67$  & 0.21  & $21.67$  & 0.21  \\
$\mu_1$ [mas yr$^{-1}$] & $0.428$  & 0.36  & $0.428$  & 0.36  \\
$\mu_2$ [mas yr$^{-1}$] & $-0.902$ & 0.35  & $-0.902$ & 0.35  \\
$v_{\rm los}$ [km s$^{-1}$] & $-104.53$ & 1.82 & $-104.53$ & 1.47 \\
\hline
\multicolumn{5}{l}{Test 2: modified $\mu_2$} \\
$\phi_1$ [deg]        & $-0.533$ & --    & $-0.533$ & --    \\
$\phi_2$ [deg]        & $0.649$  & 1.86  & $0.649$  & 1.24  \\
$d_\odot$ [kpc]       & $21.67$  & 0.20  & $21.67$  & 0.21  \\
$\mu_1$ [mas yr$^{-1}$] & $0.428$  & 0.36  & $0.428$  & 0.36  \\
$\mu_2$ [mas yr$^{-1}$] & $-0.732$ & 1.08  & $-0.732$ & 1.07  \\
$v_{\rm los}$ [km s$^{-1}$] & $-104.53$ & 1.40 & $-104.53$ & 1.78 \\
\hline
\multicolumn{5}{l}{Test 3: modified $d_\odot$} \\
$\phi_1$ [deg]        & $-0.533$ & --    & $-0.533$ & --    \\
$\phi_2$ [deg]        & $0.649$  & 2.17  & $0.649$  & 1.05  \\
$d_\odot$ [kpc]       & $17.50$  & 11.25 & $17.50$  & 10.90 \\
$\mu_1$ [mas yr$^{-1}$] & $0.428$  & 0.47  & $0.428$  & 0.42  \\
$\mu_2$ [mas yr$^{-1}$] & $-0.902$ & 0.35  & $-0.902$ & 0.34  \\
$v_{\rm los}$ [km s$^{-1}$] & $-104.53$ & 2.14 & $-104.53$ & 1.12 \\
\hline
\multicolumn{5}{l}{Test 4: compromise in $d_\odot$ and $\mu_2$} \\
$\phi_1$ [deg]        & $-0.533$ & --    & $-0.533$ & --    \\
$\phi_2$ [deg]        & $0.649$  & 1.30  & $0.649$  & 1.17  \\
$d_\odot$ [kpc]       & $20.60$  & 0.71  & $20.60$  & 0.74  \\
$\mu_1$ [mas yr$^{-1}$] & $0.428$  & 0.37  & $0.427$  & 0.36  \\
$\mu_2$ [mas yr$^{-1}$] & $-0.742$ & 0.99  & $-0.767$ & 0.80  \\
$v_{\rm los}$ [km s$^{-1}$] & $-104.53$ & 1.47 & $-104.53$ & 1.52 \\
\hline
\multicolumn{5}{l}{Test 5: modified $v_{\rm los}$} \\
$\phi_1$ [deg]        & $-0.533$ & --    & $-0.533$ & --    \\
$\phi_2$ [deg]        & $0.649$  & 1.29  & $0.649$  & 1.15  \\
$d_\odot$ [kpc]       & $20.60$  & 0.71  & $20.60$  & 0.74  \\
$\mu_1$ [mas yr$^{-1}$] & $0.428$  & 0.37  & $0.427$  & 0.36  \\
$\mu_2$ [mas yr$^{-1}$] & $-0.742$ & 0.99  & $-0.767$ & 0.80  \\
$v_{\rm los}$ [km s$^{-1}$] & $-106.90$ & 1.12 & $-108.53$ & 0.80 \\
\hline
\end{tabular}
\tablefoot{
The adopted phase-space points are denoted IC in the table.
}
\end{table}

\end{appendix}

\end{document}